\documentclass[runningheads]{llncs}
\usepackage{graphicx,verbatim}
\usepackage{amsmath}
\usepackage{bbm}
\usepackage{amsmath}
\usepackage{multirow}
\usepackage{amssymb}
\usepackage{bm}
\usepackage{pdfpages}
\usepackage{siunitx}
\usepackage{booktabs}
\usepackage{color,soul}
\usepackage[T1]{fontenc}
\usepackage{booktabs}
\usepackage{hyperref}
\usepackage{marvosym}
\usepackage{cite} 
\newcommand{\Dice}{\mathcal{D}}
\newcommand{\CE}{\mathcal{H}}

\newcommand{\fEMA}{f^{\text{ema}}}
\newcommand{\Mix}{\operatorname{Mix}}

\usepackage[]{acronym}
\acrodef{DL}{Deep Learning}
\acrodef{US}{Ultrasound}
\acrodef{PCCL}{{\bf P}ixel-level and {\bf C}lass-level {\bf C}onsistency {\bf L}earning}
\acrodef{CNN}{Convolutional Neural Network}
\acrodef{ROI}{Region of Interest}
\acrodef{SSL}{Semi-Supervised Learning}
\acrodef{GT}{Ground Truth}
\acrodef{CR}{Consistency Regularization}
\acrodef{EMA}{Exponential Moving Average}
\acrodef{MAC}{Mutual Agreement Consistency}
\acrodef{KL}{Kullback–Leibler divergence}
\acrodef{MIG}{Mutual Information Gap}
\acrodef{CE}{Cross-Entropy}
\acrodef{Dice}{Dice based coefficient}
\acrodef{DACL}{Dual Agreement Consistency Learning}
\acrodef{DAC}{Dual-Agreement Consistency}
\acrodef{SOTA}{state-of-the-art}
\acrodef{CPS}{cross pseudo supervision}
\acrodef{FS}{fully supervised}

\begin{document}
\title{Dual Agreement Consistency Learning for Semi-Supervised Fetal Ultrasound Segmentation}
\titlerunning{DACL for Semi-Supervised Ultrasound Segmentation}
%
\author{Fangyijie Wang\inst{1,2,\textsuperscript{\Letter}}\orcidID{0009-0003-0427-368X} \and
Gu\'enol\'e Silvestre\inst{1,3} \and
Ziyang Wang\inst{4} \and
Kathleen M. Curran\inst{1,2}\orcidID{0000-0003-0095-9337}}
\authorrunning{F. Wang et al.}
%
\institute{
Research Ireland Centre for Research Training in Machine Learning \and
School of Medicine, University College Dublin, Dublin, Ireland
\email{fangyijie.wang@ucdconnect.ie} \and
School of Computer Science, University College Dublin, Dublin, Ireland \and
School of Computer Science and Digital Technologies, Aston University, Birmingham, UK
}

  
\maketitle              
\begin{abstract}


Maternal-fetal \ac{US} is the primary imaging m-\\odality for monitoring fetal development, yet accurate automated segmentation remains challenging due to the scarcity of pixel-level annotations. To address this issue, we propose \ac{DACL}, a semi-supervised framework for robust fetal \ac{US} image segmentation. \ac{DACL} jointly trains a deployment-oriented lightweight convolutional network ($1.47\,\text{M}$ parameters) and a Transformer-based network, leveraging labeled data for supervised learning and unlabeled data via \ac{CPS}. 
To enhance prediction stability, we introduce a dual-agreement consistency loss that couples pixel-wise probabilistic divergence with entropy-guided confidence alignment. Unlike conventional \ac{CPS} methods that enforce agreement only at the prediction level, \ac{DACL} explicitly regularizes both distributional alignment and uncertainty, thereby suppressing unreliable pseudo-labels and enabling stable cross-architecture pseudo-label learning under extreme annotation scarcity.
Furthermore, an interpolation-based consistency strategy using mixup is applied to unlabeled samples to enhance robustness. 
Under 5\% labeled data, \ac{DACL} improves Dice by up to 2.77\% and reduces HD95 by up to 14.69 mm compared with the strongest recent semi-supervised methods, demonstrating significant improvements in boundary accuracy on both fetal head and abdomen datasets.
These results demonstrate the effectiveness of agreement-based consistency learning for annotation-efficient fetal \ac{US} segmentation. Our code is on \href{https://github.com/13204942/DACL-for-Semi-supervised-Ultrasound-Segmentation}{GitHub}.



\end{abstract}
\section{Introduction}
\label{sec:intro}

\acf{US} imaging is widely used for prenatal evaluation of fetal growth, fetal anatomy, gestational age estimation, and pregnancy monitoring due to its portability, low cost, and non-invasive nature~\cite{Salomon:2011}.
Accurate measurements enable precise evaluation of fetal biometry and effective monitoring of fetal growth~\cite{Espinoza:2013,Papageorghiou:2014}. Therefore, precise delineation of fetal structures of interest, such as the head and abdomen, is crucial for obstetricians. However, this process is patient-specific, operator-dependent, and prone to intra-\nobreakdash and inter-operator variability, which can introduce errors in fetal biometry assessment~\cite{Sarris:2012,Espinoza:2013}. Such errors may lead to inaccurate evaluation of fetal development and health, potentially resulting in missed detection of congenital diseases~\cite{Mongelli:1998}.

Recent advancements in \ac{DL} have notably improved fetal \ac{US} image segmentation. Several studies have explored lightweight \ac{DL} models for \ac{US} segmentation~\cite{Li:2025}. Nevertheless, collecting large numbers of annotated \ac{US} images for training remains labor-intensive and time-consuming, requiring medical proficiency and clinical expertise for precise pixel-level labeling~\cite{Zegarra:2023}. Recent studies~\cite{Jiang:2024,Lyu:2025} have also explored \ac{SSL} for fetal \ac{US} image analysis; however, this line of work remains relatively underexplored. Developing lightweight models with \ac{US} is particularly challenging when analyzing fetal \ac{US} images under limited annotation settings.

To address this challenge, we propose \acf{DACL}, a semi-supervised framework that enforces prediction agreement between heterogeneous models at the pixel level. \ac{DACL} adopts a cross-teaching strategy involving a lightweight \ac{CNN}, a Transformer network, and an \ac{EMA} Transformer teacher. The \ac{CNN} and Transformer students are independently supervised using \ac{GT} annotations on labeled data, while unlabeled data are leveraged through \ac{CPS}. To further improve training stability, we introduce a \ac{DAC} loss that encourages both pixel-wise agreement and uncertainty alignment between the \ac{CNN} and Transformer predictions. 
In addition, we apply an interpolation-based consistency regularization strategy (mixup) to unlabeled data, enabling the teacher model to guide the Transformer student with consistent targets. Together, these components enhance the robustness of the lightweight \ac{CNN} under limited annotation settings.

The main contributions of this paper are as follows:
(1) We propose \acf{DACL}, a semi-supervised \ac{SSL} framework that stabilizes cross pseudo supervision through a novel pixel-wise dual-agreement consistency strategy, jointly enforcing probabilistic alignment and confidence-aware regularization between heterogeneous models.
(2) By explicitly suppressing unreliable pseudo-labels, \ac{DACL} effectively leverages unlabeled data to improve fetal \ac{US} segmentation, particularly enhancing boundary accuracy under limited annotation settings.
(3) Extensive experiments on two fetal head and abdomen datasets demonstrate consistent improvements over recent semi-supervised methods across overlap and boundary metrics. 

\begin{figure}[t]
\begin{center}
\includegraphics[width=\textwidth]{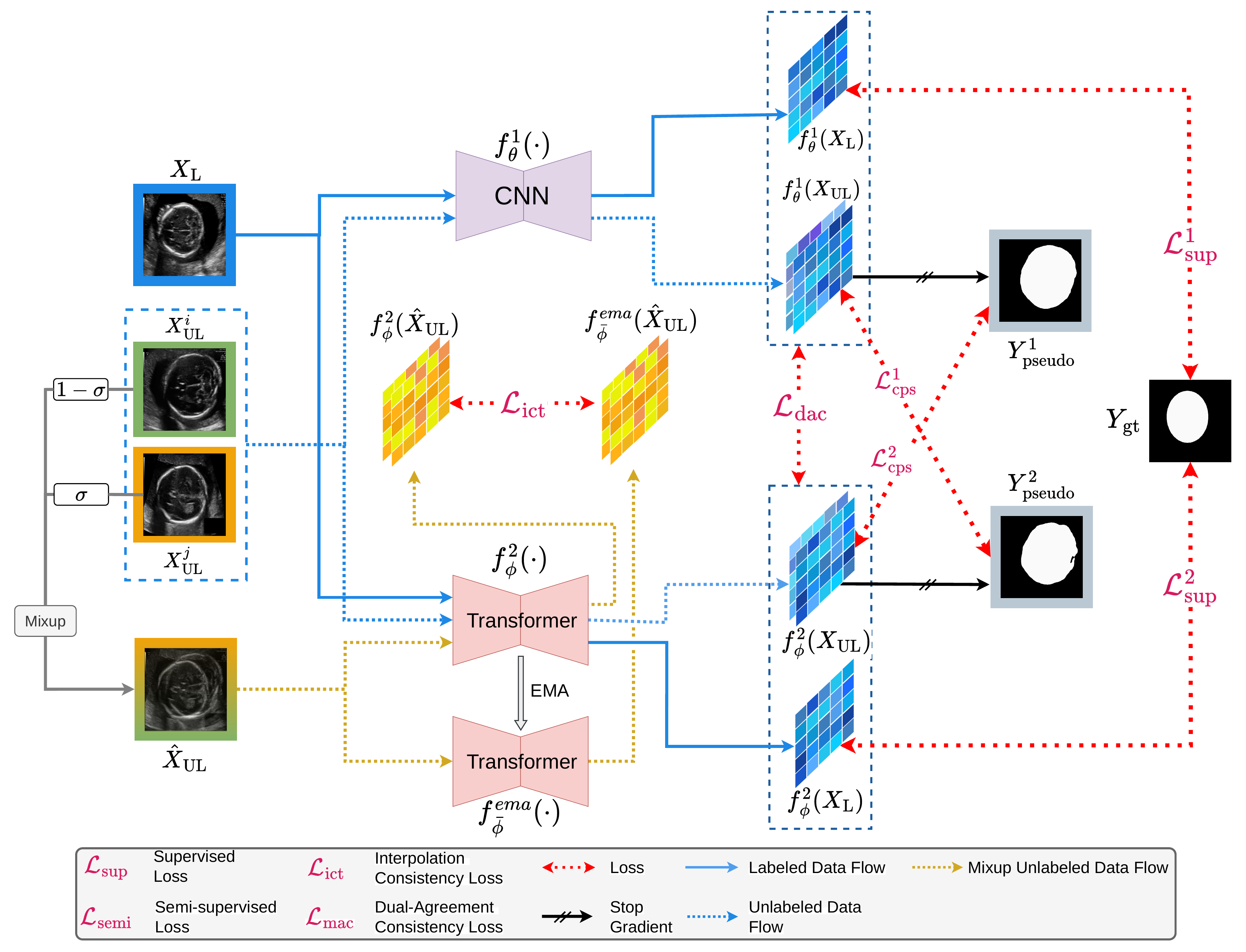}
\end{center}
\label{fig:DACL}
\caption{Overview of the proposed Dual Agreement Consistency Learning (DACL) framework. A lightweight CNN and a Transformer are jointly trained using supervised learning on labeled data and cross pseudo supervision on unlabeled data. A dual-agreement consistency loss enforces pixel-wise distribution alignment and uncertainty-aware agreement between their predictions, while mixup-based interpolation consistency with an EMA teacher further stabilizes training.}
\end{figure}

\section{Methodology}
\label{sec:Method}

This section reviews Swin-Unet~\cite{Cao:2023} and UNeXt~\cite{Valanarasu:2022}. Subsequently, we explain our semi-supervised framework \ac{DACL} in the following sections~\ref{ssec:cotrain},~\ref{ssec:ict},~\ref{ssec:dac}, and~\ref{ssec:learningloss}. The overview of our method is illustrated in Fig.\ref{fig:DACL}.


We follow the original UNeXt~\cite{Valanarasu:2022} design to build our lightweight model $f^1_{\theta}(\cdot)$. The number of channels in the encoder path is set to $\{32,64,128,160,256\}$ to ensure computational efficiency.
The $f^2_{\phi}(\cdot)$ model is implemented using the tiny Swin-Unet architecture introduced in the original work~\cite{Cao:2023}. To match the resolution of our input images, we set the shifted window size in Swin-Unet to 7. In the encoder path of Swin-Unet, the feature dimensions of each block are [$96, 192, 384$], while the bottleneck feature dimension is 768.
In terms of computational efficiency, Swin-Unet uses $27.15\text{M}$ parameters and 71.17 GFLOPs, representing a high-capacity model with strong representation ability. In contrast, UNeXt serves as our target lightweight model, requiring only $1.47\text{M}$ parameters and 7.03 GFLOPs. This substantial gap in model complexity motivates our semi-supervised framework, which aims to transfer knowledge from the powerful Transformer-based model to the lightweight model, enabling accurate yet computationally efficient fetal \ac{US} image analysis. 


\subsection{Cross-supervision between \ac{CNN} and Transformer}
\label{ssec:cotrain}

Inspired by existing cross-supervision methods, including Deep Co-Training~\cite{Qiao:2018} and \ac{CPS}~\cite{Chen:2021}, \ac{DACL} adopts a similar cross-teaching strategy that couples a lightweight \ac{CNN} with a Transformer-based network.
Given the unlabeled training dataset $\boldsymbol{X}_{\text{UL}}$, the proposed \ac{DACL} produces two predictions: $f^1_{\theta}(\boldsymbol{X}_{\text{UL}})$ and $f^2_{\phi}(\boldsymbol{X}_{\text{UL}})$.
Based on these predictions, pseudo labels for \ac{CPS} are generated as
$
\tilde{Y}^1=f_{\text{OH}}(f^1_{\theta}(\boldsymbol{X}_{\text{UL}}))
$
and 
$
\tilde{Y}^2=f_{\text{OH}}(f^2_{\phi}(\boldsymbol{X}_{\text{UL}}))
$.
No mini-batch gradient back-propagation is performed between $f^1_{\theta}(\boldsymbol{X}_{\text{UL}})$ and $\tilde{Y}^1$, or between $f^2_{\phi}(\boldsymbol{X}_{\text{UL}})$ and $\tilde{Y}^2$. Here, $f_{\text{OH}}(\boldsymbol{x}\!=\!\text{c}) = \mathbbm{1}_{[\boldsymbol{x} = \text{c}]}$ denotes the one-hot encoding function used to generate pseudo labels.
The semi-supervised loss, $\mathcal L_\text{cps}$, is the summation term $\mathcal{L}_{\text {cps}}^1 + \mathcal{L}_{\text {cps}}^2$ where
\begin{equation}
\label{semi_loss}
\begin{split}
\mathcal{L}_{\text {cps}}^1=\CE\!\left(f^1_{\theta}\left(\boldsymbol{X}_\text{UL}\right), \bm \tilde{Y}^2\right) + 
\Dice\!\left(f^1_{\theta}\left(\boldsymbol{X}_\text{UL}\right), \bm \tilde{Y}^2\right), \\
\mathcal{L}_{\text {cps}}^2=\CE\!\left(f^2_{\phi}\left(\boldsymbol{X}_\text{UL}\right), \bm \tilde{Y}^1\right) +
\Dice\!\left(f^2_{\phi}\left(\boldsymbol{X}_\text{UL}\right), \bm \tilde{Y}^1\right).
\end{split}
\end{equation}
 
\noindent $\CE(\cdot)$ and $\Dice(\cdot)$ denote the \ac{CE} loss and the standard \ac{Dice} loss, respectively. Within our framework, the Transformer-based network $f^2_{\phi}$ is used solely during training and is not used for inference.

\subsection{Interpolation Consistency Learning}
\label{ssec:ict}

To further improve the utilization of unlabeled data and the generalization of the Transformer model $f^2_{\phi}(\cdot)$, we encourage the predictions for interpolated unlabeled pixels to be consistent with the interpolation of the corresponding predictions~\cite{Verma:2022}. We adopt a mean-teacher model $\fEMA_{\bar{\phi}}(\cdot)$, where the parameters $\bar{\phi}$ are an \ac{EMA} of the student parameters $\phi$. During training, $\phi$ is updated to enforce consistency between $f^2_{\phi}(\hat{\boldsymbol{X}}_{\text{UL}})$ and $\Mix(\fEMA_{\bar{\phi}}(\boldsymbol{X}^i_{\text{UL}}), \fEMA_{\bar{\phi}}(\boldsymbol{X}^j_{\text{UL}}))$, where $\hat{\boldsymbol{X}}_{\text{UL}} = \Mix(\boldsymbol{X}^i_{\text{UL}}, \boldsymbol{X}^j_{\text{UL}})$. Here, $\Mix$ denotes the mixup operation with combination ratio $\sigma = 0.5$, and $i+j$ equals the unlabeled batch size $\text{B}_{\text{UL}}$. The consistency loss $\mathcal{L}_{\text{ict}}$ is defined as $
\label{con_loss}
\mathcal{L}_\text{ict} = \sum(f^2_{\phi}(\hat{\boldsymbol{X}}_\text{UL}) - \mathrm{Mix}(\fEMA_{\bar{\phi}}(\boldsymbol{X}^i_\text{UL}), \fEMA_{\bar{\phi}}(\boldsymbol{X}^j_\text{UL})))^2
$.

\subsection{Dual-Agreement Consistency}
\label{ssec:dac} 
To improve robustness in semi-supervised segmentation, we introduce the \ac{DAC} loss $\mathcal{L}_{\text{dac}}$. This loss enforces prediction agreement between two collaborative models: a lightweight \ac{CNN} $f^1_{\theta}$ and a Transformer $f^2_{\phi}$. The proposed $\mathcal{L}_{\text{dac}}$ consists of two complementary components: a pixel-wise \ac{KL} loss~\cite{Joyce:2021}, $\mathcal{L}_{\text{KL}}$, and an entropy-based agreement regularization term, $\mathcal{L}_{\text{ent}}$. Specifically, $\mathcal{L}_{\text{KL}}$ aligns the predicted class distributions, while $\mathcal{L}_{\text{ent}}$ encourages consistent and confident predictions between the two models.
The proposed $\mathcal{L}_{\text{dac}}$ is defined as:
$\mathcal{L}_{\mathrm{dac}}=\mathcal{L}_{\mathrm{KL}}+\mathcal{L}_\text{ent}$. $\mathcal{L}_\text{KL}$ and $\mathcal{L}_{\mathrm{ent}}$ are defined as:
\[
\label{dac_loss}
\begin{split}
\mathcal{L}_{\mathrm{KL}}&=\mathbb{E}\left[\frac{1}{H W} \sum_u \operatorname{KL}(p(u) \| q(u))\right] \\
\mathcal{L}_\text{ent}&=\mathbb{E}\left[\frac{1}{H W} \sum_u\left(H(p(u))+H(q(u))-H\left(\frac{p(u)+q(u)}{2}\right)\right)\right]
\end{split}
\]
where $u$ is a spatial pixel index $(h,w)$, and $p(\cdot)$ and $q(\cdot)$ denote the predicted class distributions at pixel $u$ from $f^1_{\theta}$ and $f^2_{\phi}$, respectively. Here, $H(\cdot)$ denotes the Shannon entropy of a categorical distribution~\cite{Shannon:1948}. 
Note that the proposed entropy-based agreement loss operates at the pixel level and is designed as an empirical regularizer rather than a formal mutual information estimator.



\subsection{The Overall Objective Function}
\label{ssec:learningloss}

All training losses are indicated by red dashed lines in Fig.~\ref{fig:DACL}. The overall training objective is a joint loss with four components: a supervised loss, $\mathcal{L}_{\text{sup}}$; a cross-supervision loss, $\mathcal{L}_{\text{cps}}$; an EMA-based interpolation consistency loss, $\mathcal{L}_{\text{ict}}$; and a dual-agreement consistency loss, $\mathcal{L}_{\text{dac}}$.
The total loss is:
\begin{equation}
\label{loss}
\mathcal{L}_{\text {total}}=(\mathcal{L}_{\text {sup }}^1+\mathcal{L}_{\text {sup }}^2)+\lambda(\mathcal{L}_{\text {cps}}^1+\mathcal{L}_{\text {cps}}^2)+\tau \mathcal{L}_{\text {ict}} + \beta \mathcal{L}_{\text {dac}} \nonumber
\end{equation}
where $\{\lambda$, $\tau$, $\beta\}$ are linear trade-off hyper-parameters set to $\{2.0, 2.0, 10.0\}$, respectively.
$\mathcal{L}^1_\text{sup}$ and $\mathcal{L}^2_\text{sup}$ are the supervision losses for $f^1_{\theta}(\cdot)$ and $f^2_{\phi}(\cdot)$ based on the labeled data $\bm{X}_\text{L}$. They are designed with a combination of the \ac{Dice} and \ac{CE} losses, as follows:
\[
\begin{split}
\mathcal{L}_{\text {sup }}^1 &=\CE\!\left(f^1_{\theta}\left(\bm{X}_\text{L}\right), \bm{Y}_{\mathrm{gt}}\right)+\Dice\!\left(f^1_{\theta}\left(\bm{X}_\text{L}\right), \bm{Y}_{\mathrm{gt}}\right) \\
\mathcal{L}_{\text {sup }}^2&=\CE\!\left(f^2_{\phi}\left(\bm{X}_\text{L}\right), \bm{Y}_{\mathrm{gt}}\right)+\Dice\!\left(f^2_{\phi}\left(\bm{X}_\text{L}\right), \bm{Y}_{\mathrm{gt}}\right)
\end{split}
\]
cross-supervision losses $\mathcal{L}_{\text {cps}}^1$ and $\mathcal{L}_{\text {cps}}^2$ are given in (\ref{semi_loss}). The loss $\mathcal{L}_{\text {ict}}$ and $\mathcal{L}_{\text {dac}}$ are defined in Section~\ref{con_loss} and Section~\ref{dac_loss}, respectively.

\section{Experiments and Results}
\label{sec:ex_result}

\subsection{Datasets}
\label{ssec:data}

We used two public datasets in this study: the \textbf{HC18} dataset, collected using General Electric \ac{US} devices in the Netherlands~\cite{Heuvel:2018_b}, and the \textbf{F-Abd} dataset, acquired by novice users using a low-cost portable probe connected to a smartphone in low-income countries~\cite{Sappia:2025}.
The HC18 dataset contains fetal head standard planes annotated for head circumference measurement, whereas the F-Abd dataset contains images annotated with optimal planes for abdominal circumference measurement.
To use the pre-trained Transformer models $f^2_{\phi}$ and $\fEMA_{\bar{\phi}}$ in our framework, we convert all \ac{US} images to RGB format. 

\textbf{HC18:} All images display a resolution of $800 \times 540$ pixels. It contains 500 training images (450 subjects), 50 validation images (50 subjects), and 449 test images (410 subjects). \textbf{F-Abd:} Each image has a resolution of $744 \times 562$ pixels. It includes 1084 training images (94 subjects), 139 validation images (18 subjects), and 922 test images (74 subjects). 

\subsection{Implementation Details}

We built \ac{DACL} by employing two segmentation model architectures, Swin-Unet~\cite{Cao:2023} and UNeXt~\cite{Valanarasu:2022}. For the Swin-Unet model, we used pre-trained weights \cite{Cao:2023} to initialize it. We trained \ac{DACL} for 400 epochs, with a labeled batch size of 1 and an unlabeled batch size of 4. A stochastic gradient descent optimizer was used with an initial learning rate of 0.01 and a momentum value of 0.9. The weight decay was set to 0.0001. Our code is implemented in Python (3.11.5) with PyTorch (2.1.2) and CUDA (12.2) on a single NVIDIA 4090 GPU. The model was evaluated on the validation set each epoch, and the best-performing UNeXt weights are saved. The above setting is directly applied to all other baseline methods without any modification.

\begin{table}[bp]
\centering
\caption{The quantitative results on the HC18 dataset. The best results are in {\bf bold}. The $2^\text{nd}$ best results are in \underline{underline}. 
}
\label{hc18_res}
\scalebox{0.8}{
\begin{tabular}{llccccccc}
\hline
& \multirow{2}{*}{Method} & \# & \# & \multirow{2}{*}{DSC $\uparrow$} & \multirow{2}{*}{Jaccard $\uparrow$} & \multirow{2}{*}{HD95 $\downarrow$} & \multirow{2}{*}{ASD $\downarrow$} & \multirow{2}{*}{$p$-value}\\
& & Labeled & Unlabeled &  &  &  &  & \\
\hline
\multirow{2}{*}{FS} & UNeXt~\cite{Valanarasu:2022} & 500  & \multirow{2}{*}{-} & 93.35(8.58) & 88.51(12.36) & 39.11(41.87) & 13.42(15.45) & 0.000 \\
& Swin-Unet~\cite{Cao:2023} & (100\%) & & \textbf{96.96(2.88)} & \textbf{94.23(4.70)} & \textbf{9.54(11.91)} & \textbf{3.89(4.08)} & 0.000 \\
\hline\hline
\multirow{18}{*}{SSL} & MT (2017)~\cite{Tarvainen:2017} & \multirow{9}{*}{25} & \multirow{9}{*}{475} & 89.53(10.41) & 82.37(14.22) & 54.79(36.98) & 18.90(15.26) & 0.000 \\
& DCT (2018)~\cite{Qiao:2018} & \multirow{9}{*}{(5\%)} & \multirow{9}{*}{(95\%)} & 91.69(10.01) & 85.92(13.76) & 41.62(36.06) & 14.57(14.69) & 0.000 \\
& CPS (2021)~\cite{Chen:2021} & & & 91.40(10.10) & 85.42(13.70) & 44.88(36.84) & 15.59(15.10) & 0.000 \\
& ICT (2022)~\cite{Verma:2022} & & & 91.70(10.21) & 85.97(13.87) & 46.06(40.56) & 15.80(15.77) & 0.000 \\
& CTCT (2022)~\cite{Luo:2022} & & & 91.57(9.84) & 85.66(13.45) & 43.45(36.00) & 14.94(14.74) & 0.000 \\
& PCPCS (2024)~\cite{Ma:2024}  & & & 92.30(9.07) & 86.76(12.69) & 41.13(37.76) & 13.88(14.11) & 0.000 \\
& DSTCT (2024)~\cite{Jiang:2024}  & & & 93.97(7.78) & \underline{89.44(11.14)} & \underline{26.49(30.86)} & \underline{9.35(12.14)} & 0.000 \\
& LMCT (2025)~\cite{Wang:2025b}  & & & \underline{93.98(7.43)} & 89.39(10.80) & 27.01(30.82) & 9.49(11.96) & 0.016 \\
& \bf DACL (Ours) & & & \textbf{95.05(5.41)} & \textbf{90.99(8.28)} & \textbf{22.54(28.02)} & \textbf{7.86(9.37)} & - \\
\cline{2-9}
& MT (2017)~\cite{Tarvainen:2017} & \multirow{9}{*}{50} & \multirow{9}{*}{450} & 94.43(7.20) & 90.15(10.53) & 29.59(37.07) & 10.08(12.45) & 0.001 \\
& DCT (2018)~\cite{Qiao:2018} & \multirow{9}{*}{(10\%)} & \multirow{9}{*}{(90\%)} & 94.91(7.20) & 91.02(10.30) & 24.00(33.04) & 8.55(11.56) & 0.029 \\
& CPS (2021)~\cite{Chen:2021} & & & 94.67(7.31) & 90.61(10.51) & 26.74(34.33) & 9.18(12.02) & 0.006 \\		
& ICT (2022)~\cite{Verma:2022} & & & 94.76(6.94) & 90.70(10.06) & 26.52(34.84) & 9.42(12.40) & 0.008 \\
& CTCT (2022)~\cite{Luo:2022} & & & 95.15(6.35) & 91.31(9.32) & 22.60(29.93) & 7.99(10.50) & 0.084 \\
& PCPCS (2024)~\cite{Ma:2024}  & & & 94.26(7.90) & 89.97(11.14) & 27.26(34.46) & 9.81(13.18) & 0.000 \\
& DSTCT (2024)~\cite{Jiang:2024}  & & & 95.17(6.42) & 91.35(9.39) & 20.37(27.15) & 7.27(9.40) & 0.097 \\
& LMCT (2025)~\cite{Wang:2025b}  & & & \underline{95.76(4.88)} & \textbf{92.21(7.51)} & \underline{19.45(25.81)} & \underline{6.78(8.17)} & 0.974 \\
& \bf DACL (Ours) & & & \textbf{95.77(4.16)} & \underline{92.15(6.66)} & \textbf{19.42(25.29)} & \textbf{6.75(7.86)} & - \\
\hline
\end{tabular}
}
\end{table}

\subsubsection{Data Augmentation:} In this study, data augmentations are applied to the labeled training data $\bm X_\text{L}$. 
The training and testing images are resized to $448 \times 448$ to facilitate computational resource demands. The detailed parameters of the data augmentation techniques are in our code repository on \href{https://github.com/13204942/DACL-for-Semi-supervised-Ultrasound-Segmentation}{GitHub}.

\subsubsection{Evaluation Metrics:} A range of quantitative metrics are used for testing methods, including the Dice Similarity Coefficient (DSC (\%)), Jaccard (\%), Average Surface Distance (ASD (mm)), and 95\% Hausdorff Distance (HD95 (mm)). For fair comparison, all SSL frameworks use two networks: UNeXt and Swin-Unet for CNN-Transformer co-training, and two UNeXt networks otherwise.

\subsection{Results}

\begin{table}[tp]
\centering
\caption{The quantitative results on the F-Abd dataset. The best results are in {\bf bold}. The $2^\text{nd}$ best results are in \underline{underline}. 
}
\label{fabd}
\scalebox{0.77}{
\begin{tabular}{llccccccc}
\hline
& \multirow{2}{*}{Method} & \# & \# & \multirow{2}{*}{DSC $\uparrow$} & \multirow{2}{*}{Jaccard $\uparrow$} & \multirow{2}{*}{HD95 $\downarrow$} & \multirow{2}{*}{ASD $\downarrow$} & \multirow{2}{*}{$p$-value}\\
& & Labeled & Unlabeled &  &  &  &  & \\
\hline\hline
\multirow{2}{*}{FS} & UNeXt & 1084 & \multirow{2}{*}{-} & 88.74(7.11) & 80.42(10.47) & 44.87(42.34) & 14.94(13.20) & 0.000 \\
& Swin-Unet & (100\%) & & \textbf{93.24(3.79)} & \textbf{87.55(6.23)} & \textbf{15.64(19.05)} & \textbf{6.24(7.30)} & 0.000 \\
\hline
\multirow{18}{*}{SSL} & MT (2017)~\cite{Tarvainen:2017} & \multirow{9}{*}{29} & \multirow{9}{*}{1055} & 61.78(13.69) & 46.10(14.26) & 112.61(46.91) & 46.12(19.00) & 0.000 \\
& DCT (2018)~\cite{Qiao:2018} & \multirow{9}{*}{(5\%)} & \multirow{9}{*}{(95\%)} & 61.22(15.89) & 45.98(16.45) & 113.36(42.26) & 47.17(17.78) & 0.000 \\
& CPS (2021)~\cite{Chen:2021} & & & \underline{64.43(12.92)} & 48.87(14.14) & 101.61(38.44) & 43.05(16.06) & 0.000 \\
& ICT (2022)~\cite{Verma:2022} & & & 63.14(15.02) & 47.87(15.90) & 125.01(52.69) & 50.59(21.34) & 0.000 \\
& CTCT (2022)~\cite{Luo:2022} & & & 63.77(15.90) & 48.80(17.11) & 100.05(42.00) & 42.62(18.81) & 0.000 \\
& PCPCS (2024)~\cite{Ma:2024}  & & & 63.64(17.49) & \underline{49.02(18.51)} & \underline{90.13(42.45)} & \underline{38.96(18.21)} & 0.000 \\
& DSTCT (2024)~\cite{Jiang:2024}  & & & 58.89(13.85) & 43.09(13.85) & 90.86(30.69) & 43.11(15.41) & 0.000 \\
& LMCT (2025)~\cite{Wang:2025b}  & & & 62.06(13.41) & 46.34(13.98) & 99.34(41.52) & 42.25(17.22) & 0.000 \\
& \bf DACL (Ours) & & & \textbf{66.53(13.67)} & \textbf{51.43(15.54)} & \textbf{74.43(32.17)} & \textbf{35.14(15.22)} & -\\
\cline{2-9}
& MT (2017)~\cite{Tarvainen:2017} & \multirow{9}{*}{93} & \multirow{9}{*}{991} & 73.51(21.63) & 61.76(21.71) & \underline{58.42(36.44)} & \underline{23.07(15.05)} & 0.050 \\
& DCT (2018)~\cite{Qiao:2018} & \multirow{9}{*}{(10\%)} & \multirow{9}{*}{(90\%)} & 73.43(18.40) & 60.73(18.97) & 59.17(32.33) & 23.95(14.19) & 0.021 \\
& CPS (2021)~\cite{Chen:2021} & & & \textbf{76.32(15.53)} & \textbf{63.93(18.06)} & 75.38(42.83) & 29.19(18.06) & 0.081 \\		
& ICT (2022)~\cite{Verma:2022} & & & 71.59(20.30) & 58.96(20.59) & 72.40(34.64) & 29.81(16.01) & 0.000 \\
& CTCT (2022)~\cite{Luo:2022} & & & 72.88(13.92) & 59.19(17.05) & 75.89(37.19) & 31.93(16.94) & 0.000 \\
& PCPCS (2024)~\cite{Ma:2024}  & & & 70.61(14.53) & 56.48(17.13) & 82.46(37.03) & 35.17(17.31) & 0.000 \\
& DSTCT (2024)~\cite{Jiang:2024}  & & & 64.63(19.24) & 50.45(19.32) & 61.58(31.85) & 29.28(15.18) & 0.000 \\
& LMCT (2025)~\cite{Wang:2025b}  & & & 73.21(17.36) & 60.21(18.32) & 73.43(40.92) & 29.20(18.23) & 0.002 \\
& \bf DACL (Ours) & & & \underline{75.15(12.98)} & \underline{61.84(15.90)} & \textbf{49.49(27.48)} & \textbf{21.64(11.93)} & - \\
\hline
\end{tabular}
}
\end{table}

\subsubsection{Quantitative Results:} 
We compared \ac{DACL} with several \ac{SSL} baselines using 5\% and 10\% of the labeled training data. Table~\ref{hc18_res} shows that \ac{DACL} achieves competitive performance across the compared methods. Notably, \ac{DACL} consistently improves the performance of the lightweight UNeXt model and reveals that existing \ac{SSL} frameworks remain less effective for lightweight segmentation networks under annotation scarcity. In particular, relative to LMCT (a strong baseline), \ac{DACL} improves DSC by 1.07\% ($p=0.016$) on the HC18 dataset with 5\% labeled data, while reducing HD95 by 4.47, indicating improved boundary accuracy. Under the 10\% labeled setting, DACL and LMCT achieved comparable performance ($p = 0.974$), indicating that both methods achieved a similar level of performance. Additionally, \ac{DACL} outperforms the \ac{FS} UNeXt model across all metrics ($p < 0.05$), demonstrating its ability to exploit unlabeled data more effectively than \ac{FS} training.

On the more challenging F-Abd dataset (Table~\ref{fabd}), \ac{DACL} achieves competitive segmentation accuracy among semi-supervised methods and yields \ac{SOTA} performance on boundary-sensitive metrics ($p < 0.05$) with 5\% labeled data. In particular, DACL achieved statistically significant improvements over LMCT ($p < 0.05$) across all metrics. However, under the 10\% labeled setting, \ac{DACL} performs slightly below the best competing method in DSC and Jaccard. Nevertheless, a performance gap remains between semi-supervised and \ac{FS} methods. This gap can be attributed to the extremely limited number of labeled samples and the use of low-cost portable ultrasound probes, which produce noisier images and are prone to both intra- and inter-operator variability.

\begin{figure}[tbp]
\begin{center}
\includegraphics[width=\textwidth]{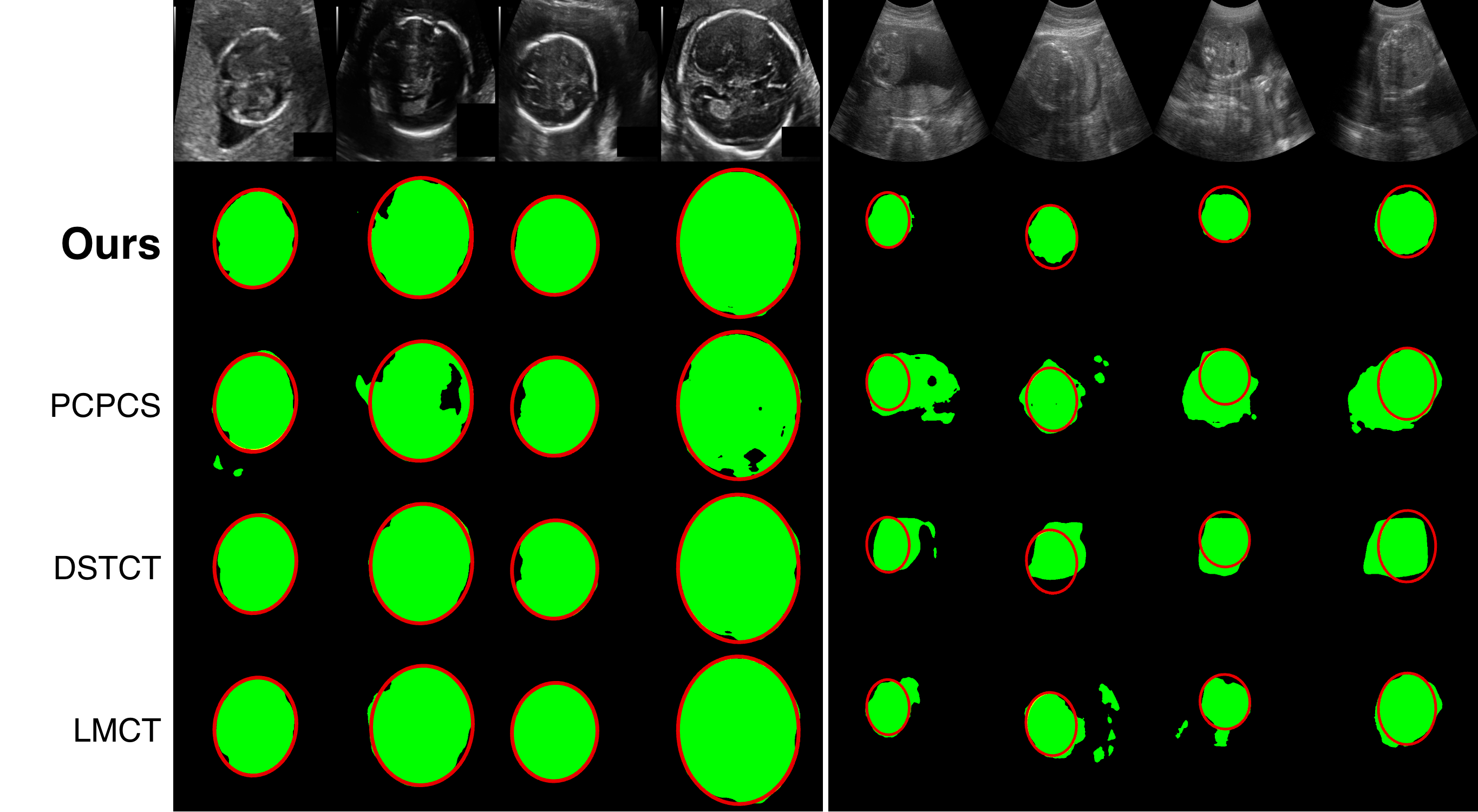}
\end{center}
\caption{Visual comparison of latest methods (from 2024) when using 10\% labeled data for testing. From left to right, they are four HC18 and four F-Abd samples.}
\label{fig:res_vis}
\end{figure}

\subsubsection{Qualitative Results:}
Fig.~\ref{fig:res_vis} compares the segmentation results of \ac{DACL} with \ac{GT} masks and the three latest semi-supervised methods. On the fetal head dataset, \ac{DACL} shows performance comparable to the competing methods, with only modest visual differences. In contrast, on the fetal abdomen dataset, \ac{DACL} generates significantly more accurate and complete abdomen \ac{ROI} delineations, with fewer missing regions and smoother boundaries, indicating a clearer advantage under the more challenging imaging conditions.

\subsubsection{Ablation Study:}
Table~\ref{ablation} presents the results of our ablation study. The second row indicates that \ac{CPS} provides effective supervision signals from unlabeled data by adding $\mathcal{L}_{\text{cps}}$ to the baseline.  Incorporating the interpolation consistency loss $\mathcal{L}_{\text{ict}}$ further boosts accuracy on both datasets, suggesting that teacher-guided mixup regularization stabilizes training and improves generalization. Interestingly, replacing $\mathcal{L}_{\text{ict}}$ with the proposed dual-agreement loss $\mathcal{L}_{\text{dac}}$ yields the largest boundary-quality gains (HD95/ASD), particularly on F-Abd, although it can trade off some overlap accuracy (DSC). 
Finally, the last row shows that combining all three losses yields the best performance on both fetal head and abdomen segmentation, confirming the overall effectiveness of our method.

\begin{table}[t]
\centering
\caption{Ablation studies for each loss function on the HC18 and F-Abd datasets with 10\% of the data labeled. The best results are in {\bf bold}. 
}
\label{ablation}
\scalebox{.8}{
\begin{tabular}{l|ccc|ccc}
\hline
\multirow{2}{*}{Method} & \multicolumn{3}{c|}{HC18} & \multicolumn{3}{c}{F-Abd} \\
& DSC $\uparrow$ & HD95 $\downarrow$ & ASD $\downarrow$ & DSC $\uparrow$ & HD95 $\downarrow$ & ASD $\downarrow$ \\
\hline
Baseline & 92.36(9.28) & 36.82(36.01) & 13.22(14.07) & 70.88(21.46) & 75.02(39.70) & 30.18(17.83) \\
+ $\mathcal{L}_{\text{cps}}$ & 93.64(7.17) & 30.48(31.94) & 10.79(11.83) & 72.24(14.85) & 72.26(36.50) & 31.08(17.01) \\
+ $\mathcal{L}_{\text{cps}}$ + $\mathcal{L}_{\text{ict}}$ & 93.84(6.90) & 30.35(31.24) & 10.55(11.11) & 74.05(17.70) & 61.83(36.70) & 26.18(17.55) \\
+ $\mathcal{L}_{\text{cps}}$ + $\mathcal{L}_{\text{dac}}$ & 94.95(6.31) & 23.44(30.29) & 8.26(10.85) & 70.47(20.33) & 52.34(38.01) & 22.08(15.46) \\
+ $\mathcal{L}_{\text{cps}}$ + $\mathcal{L}_{\text{ict}}$ + $\mathcal{L}_{\text{dac}}$ & \textbf{95.77(4.16)} & \textbf{19.42(25.29)} & \textbf{6.75(7.86)} & \textbf{75.15(12.98)} & \textbf{49.49(27.48)} & \textbf{21.64(11.93)} \\
\hline
\end{tabular}
}
\end{table}

\section{Conclusion}
We introduced \ac{DACL}, a semi-supervised framework that stabilizes cross pseudo supervision through dual-agreement consistency, jointly enforcing probabilistic agreement and uncertainty-aware confidence agreement between heterogeneous models. This strategy improves pseudo-label reliability under extreme annotation scarcity and challenging ultrasound imaging conditions while maintaining a lightweight architecture suitable for practical deployment.  Experiments on fetal head and abdomen data show consistent improvements in both overlap and boundary accuracy compared with recent semi-supervised approaches, supporting more reliable downstream fetal biometry and quantitative assessment.

\begin{credits}
\subsubsection{\ackname} This work was funded by Taighde \'{E}ireann – Research Ireland through the Research Ireland Centre for Research Training in Machine Learning \\
(18/CRT/6183).

\end{credits}

%
%
%
\bibliographystyle{splncs04}
\bibliography{paper-0556}
%




\end{document}